\documentclass[aps,aps,prb,twocolumn,floatfix,showpacs]{revtex4}
\usepackage{amsmath,amssymb,graphicx,bm}
\begin{document}
\title{Conservation laws in disordered electron systems: Thermodynamic
limit and configurational averaging}

\author{V.  Jani\v{s}} \author{ J. Koloren\v{c}}

\affiliation{Institute of Physics, Academy of Sciences of the Czech
  Republic, Na Slovance 2, CZ-18221 Praha 8, Czech Republic}
\email{janis@fzu.cz, kolorenc@fzu.cz}

\date{\today}


\begin{abstract}
We discuss conservation of probability in noniteracting disordered electron
systems. We argue that although the norm of the electron wave function is
conserved in individual realizations of the random potential, we cannot
extend this conservation law easily to configurationally averaged systems
in the thermodynamic limit. A direct generalization of the norm
conservation to averaged functions is hindered by the existence of
localized states breaking translational invariance. Such states are
elusive to the description with periodic Bloch waves. Mathematically this
difficulty is manifested through the diffusion pole in the electron-hole
irreducible vertex. The pole leads to a clash with analyticity of the
self-energy, reflecting causality of the theory, when norm conservation is
enforced by the Ward identity between one- and two-particle averaged Green
functions.
\end{abstract}

\maketitle 
\section{Introduction}\label{sec:intro}

Itinerant quantum disordered systems are very difficult to describe
quantitatively. One has to be aware of a few pitfalls that may destroy the
standard scheme adopted from clean solids. Quantum systems must be
represented in a Hilbert space. When a static, quenched disorder is
present, each single configuration of the random potential (force) has its
own state space. First assumption we standardly make is that a suitable
representation space exists comprising states of all configurations of the
random potential. The disordered system then formally behaves as a clean
one with renormalized products of operators via vertex corrections.

Operators are generally difficult to treat unless we are able to use their
spectral representations. In the thermodynamic limit, i.~e. in infinite
volume, the eigenstates of itinerant Hamiltonians may either be localized,
quadratic integrable, or extended with a plane-wave character.
The former eigenfunctions belong to the point spectrum with separate
eigenvalues, while the latter ones to a continuous band of eigenenergies.
These two types of eigenfunctions are well separable for one inhomogeneous
configuration. Configurational averaging, however, can turn the point
spectrum effectively continuous and it becomes extremely difficult to
distinguish the character of spectral energies, that is the extension of
the eigenfunctions, solely from averaged quantities. Determination of the
extension of typical eigenstates of a random itinerant Hamiltonian is the
core of the so-called Anderson localization problem.\cite{Anderson58}

Since exact solutions of the Anderson localization problem are not
available one has to resort approximate treatments. There are two
complementary tools, numerical and analytic, to tackle this problem.
Neither of them is, however, able to answer all questions about the
spectrum of random itinerant Hamiltonians. While the former technique can
deal only with finite lattices and many configurations of the random
potential,\cite{Kramer93} the latter one deals mostly with the
thermodynamic limit and configurationally averaged quantities.\cite{Lee85}
In particular the latter approach demands a special caution in building a
consistent picture of the metal-insulator transition at which
configurations with extended states become irrelevant.

In order not to miss the physics of the problem one has to minimize
spurious effects, introduced in approximate treatments, by taking
into account as many exactly valid relations as possible. Conservation
laws are typical properties to be maximally satisfied also in approximate
theories. They impose the so-called ``no-go'' theorems that may essentially
restrict the state space of the system and hint to available ``order
parameter''. The aim of this paper is to elucidate the role and
applicability of conservation of probability, or conservation of the norm
of the wave function, in the Anderson localization problem. Although it
may seem evident that the norm of the wave function must be
unconditionally conserved, it is the random character of the Anderson
model that makes the proper utilization of probability
conservation a rather subtle issue deserving a deeper and more thorough
investigation.
 
We use the simplest possible model on which we can demonstrate subtleties
of the application of conservation laws in random systems. It is the
Anderson model of noninteracting lattice electrons subject to a random
potential, the Hamiltonian of which can be written as
\begin{equation}\label{eq:AD-hamiltonian}
\widehat{H}_{AD}=\sum_{<ij>}t_{ij}c_{i}^{\dagger} c_{j}+\sum_iV_ic_{i
  }^{\dagger } c_{i} \, .
\end{equation}
We assume that the kinetic energy (nearest-neighbor hopping term) $t_{ij}=
t_{|i-j|}$ is homogeneous and only the local potential $V_i$ is random. The
values of the random potential $V_i$ are site-independent and obey a
disorder distribution $\rho(V)$. A function depending on the random
potential $V_i$ is then  averaged via
\begin{eqnarray}   \label{eq:averaging}   \left\langle
X(V_i)\right\rangle_{av}&=&\int_{-\infty}^{\infty}dV\rho(V)X(V)\, .
\end{eqnarray}

Anderson showed in his seminal paper, Ref.~\onlinecite{Anderson58}, that
the eigenstates of single configurations can either be extended or
localized. In the former case we have a metallic conductivity (diffusive
regime). The latter case, called Anderson localization, leads to an
insulator and we have a diffusionless transport regime. A consistent theory
of the disorder induced vanishing of diffusion is still missing. One of
the reasons for this shortage is incomplete understanding of probability
conservation in this simple model. We discuss in the following sections
what holds for sure and what applications of conservation laws demand
special additional assumptions.

\section{Norm conservation and Ward identity: one configuration}
\label{sec:Single-configuration}

We start with a finite system and consider one fixed configuration of the
random potential. It is a very straightforward situation where no problems
with conservation of the norm of the wave function are to be expected.
Since the Anderson model, Eq.~\eqref{eq:AD-hamiltonian}, consists of
noninteracting electrons, we can formally find an exact solution via
an evolution operator. We define
\begin{equation}\label{eq:resolvent-time}
\widehat{U}(t) =  e^{-i\widehat{H}t}\ .
\end{equation}
The evolution operator $\widehat{U}(t)$  represents a \textit{formal}
solution unless we are able to diagonalize the Anderson Hamiltonian. It
is, however, possible only numerically and for rather small lattices. We
can nevertheless use this formal solution to see how conservation of
the norm of the wave function can be represented. We evaluate the following
integral for arbitrary complex energies $z$ with $\Im z >0$
\begin{multline}\label{eq:norm-conservation}
\int_0^{\infty\!\!}dt\ e^{izt} \left\langle\psi(t)|\psi(t) \right\rangle =
\int_0^{\infty}\!\!dt\ e^{izt} \left\langle\psi\left|\widehat{U}(t)
\widehat{U}^\dagger(t)\right|\psi \right\rangle \\  = \int_0^{\infty}dt
\sum_{ijn}\psi_i^{\phantom\dagger}\psi_j^* e^{izt} \left\langle
j\left|\widehat{U}(t)\right|n\right\rangle \left\langle
n\left|\widehat{U}^\dagger(t)\right|i\right\rangle   \\ = -\frac 1{iz} \sum_n
|\psi_n|^2
\end{multline}
where  $|i\rangle$ denotes the localized orbital on the site
$\mathbf{R}_i$ and $\langle i|j\rangle = \delta_{ij}$.

Identity \eqref{eq:norm-conservation} is a rather evident consequence of
conservation of the norm of the wave function. It is less evident and more
interesting when we rewrite this identity with resolvent operators and
Green functions. We introduce the resolvent as a Fourier transform of the
evolution operator
\begin{equation}\label{eq:resolvent-operator}
\widehat{G}(z) = -i \int_{-\infty}^{\infty} dt\ \theta(t) \
e^{izt}\widehat{U}(t) = \left[z\widehat{1}-\widehat{t}
-\widehat{V}\right]^{-1}\ .
\end{equation}
Using the local orbitals (Wannier states) we define one- and two-particle
Green functions
\begin{subequations}
\begin{align}   \label{eq:resolvents-energy}
G_{ij}(z) &= \left\langle i\left|\widehat{G}(z)\right| j \right\rangle\ ,
\\
\label{eq:2p-resolvent-energy}
G^{(2)}_{ij,kl}(z_1,z_2)&= \left\langle i
\left|\widehat{G}(z_1)\right|j\right\rangle \left\langle
k\left|\widehat{G}(z_2)\right| l\right\rangle\ .
\end{align}
\end{subequations}
The two-particle Green function is trivially only a product of
one-electron Green functions, since we do not have interacting electrons
and we stay within one fixed configuration of the random potential.

It is easy to insert the resolvent operators (finite matrices) into
Eq.~\eqref{eq:norm-conservation} to obtain
\begin{multline}\label{eq:WI-finite_volume}
\sum_n G^{(2)}_{in,nj}(z_1,z_2) = \sum_n G_{in}(z_1)G_{nj}(z_2)\\ = \frac
1{z_2 - z_1} \left[G_{ij}(z_1) - G_{ij}(z_2)\right]\ .
\end{multline}
The result is kind of Ward identity being a direct consequence of
probability conservation for quantum noninteracting systems. It holds for
any configuration of the random potential.

In finite systems all eigenenergies of the Anderson Hamiltonian are
discrete and nondegenerate. The situation changes if we perform the
thermodynamic limit, where the volume of the lattice increases to
infinity, i.~e., we have infinite-many lattice sites. To perform the
thermodynamic limit we choose the values of the random potential on the
newly added sites according to the disorder distribution $\rho$ from
Eq.~\eqref{eq:averaging}. To extend the validity of the Ward
identity~\eqref{eq:WI-finite_volume} to infinite systems we have to demand
completeness of the set of Wannier states. That is, we require
\begin{align}\label{eq:completeness-Wannier}
\sum_n \left| n \right\rangle\left\langle n \right| = \widehat{1}\ , &
\quad ||\psi||^2 \equiv \sum_n \left\langle\psi\left| n \right\rangle
\left\langle n \right| \psi \right\rangle = 1\ .
\end{align}
The latter equality says that we close finite sums over the Wannier states
in such a way that the norm can be fixed. The Wannier states form an
orthonormal basis in the Hilbert space of random configurations, i.~e.
$\left\langle i | j\right\rangle =\delta_{ij}$. Condition
\eqref{eq:completeness-Wannier} guarantees validity of the Ward
identity~\eqref{eq:averaging} strictly only for proper states from the
Hilbert space (localized states). In this case the infinite sum on the
l.h.s. of Eq.~\eqref{eq:WI-finite_volume} converges strongly in the
Hilbert space spanned over the Wannier states. We, however, know that the
thermodynamic limit generally generates extended states belonging to the
continuous spectrum of the Anderson Hamiltonian. These extended states are
not proper vectors in the Hilbert space. To prove the Ward identity also
for extended states of a single configuration we have to weaken the
convergence criterion for the sum over the intermediate states. It is in
principle possible to formalize such a procedure.

\section{Configurationally averaged quantities}\label{sec:averaged}

Analysis of individual configurations of the random potential is manageable
only by numerical means, since random configurations are not generally
translationally invariant. Hence, an analytic treatment is not viable due
to the lack of apparent symmetry that would help reduce the number of
degrees of freedom.

The configurational averaging provides a means for reducing the complexity
of the description of random systems, as it restores translational
invariance. Translationally invariant lattice electrons can be formally
described by standard field-theoretic methods used for pure (interacting)
systems. If the ergodic (self-averaging) hypothesis holds, i.~e., if
extensive thermodynamic quantities evaluated for one typical configuration
equal their averaged values, averaging does not change the physics of the
problem. It does not, however, mean that all relations and identities
valid for single configurations can automatically be transferred to
averaged quantities.

To generalize the Ward identity \eqref{eq:WI-finite_volume} to averaged
quantities we have to assume that the sum over the intermediate states
commutes with the configurational averaging, that is
\begin{equation}\label{eq:averaging-commutativity}
\frac 1{\mathcal{N}_c}\sum_{\{C\}}\sum_n \Psi_n(\{C\})= \sum_n \frac
1{\mathcal{N}_c}\sum_{\{C\}} \Psi_n(\{C\})\ ,
\end{equation}
where $\Psi_n(\{C\})$ are configurationally dependent functions of
interest, the matrix elements of the two-particle Green function in this
case. This equality may seem evident and indeed holds for every
finite-volume samples as all involved sums run over finite number of
elements. In the thermodynamic limit both sums, over the basis states
$\sum_n$ and over the random configurations
$\mathcal{N}_c^{-1}\sum_{\{C\}}$, are infinite and they are
interchangeable only if states of all relevant random configurations fall
into a single Hilbert space. That is, if the basis vectors $|n\rangle$ are
configurationally independent. We will show later on that this need not be
always the case.

After averaging the system does not behave as a simple Fermi gas, since the
product of averages does not equal the average of the product. Hence,
the two-electron resolvent, Eq.~\eqref{eq:2p-resolvent-energy}, will be
defined via a Bethe-Salpeter equation,
\begin{multline}\label{eq:BS-direct}
\left\langle G^{(2)}_{ij,kl}(z_1,z_2)\right\rangle_{av} = \left\langle
G_{ij}(z_1)\right\rangle_{av} \left\langle G_{kl}(z_2)\right\rangle_{av}\\
 + \sum_{i'j'k'l'} \left\langle G_{ii'}(z_1)\right\rangle_{av}
\left\langle G_{l'l}(z_2)\right\rangle_{av}\\ \times
\Lambda_{i'j',k'l'}(z_1,z_2) \left\langle
G^{(2)}_{j'j,kk'}(z_1,z_2)\right\rangle_{av} .
\end{multline}
We introduced a two-particle irreducible vertex $\Lambda$ to take into
account correlations between the one-electron Green functions from the
product in Eq.~\eqref{eq:2p-resolvent-energy} due to the averaging.
Analogously, the averaged one-electron resolvent
\eqref{eq:resolvent-operator} is determined from a self-energy
$\widehat{\Sigma}$ via the Dyson equation, $\bigl\langle
\widehat{G}(z)\bigr\rangle_{av} =\bigl[z\widehat{1}-\widehat{t}
-\widehat{\Sigma}\bigr]^{-1}$. Both, the self-energy $\Sigma$ and the
irreducible vertex $\Lambda$, display translational invariance.

Assuming validity of Eq.~\eqref{eq:averaging-commutativity} and using the
Bethe-Salpeter equation \eqref{eq:BS-direct} we transform the Ward
identity \eqref{eq:WI-finite_volume} to an identity between the
self-energy and the two-particle irreducible vertex,\cite{Janis01b}
\begin{multline}\label{eq:WI-VW}
\Sigma_{il}(z_1) - \Sigma_{il}(z_2)\\  =
\sum_{j'k'}\Lambda_{ij',k'l}(z_1,z_2) \left[\left\langle
G_{j'k'}(z_1)\right\rangle_{av} - \left\langle
G_{j'k'}(z_2)\right\rangle_{av}\right] .
\end{multline}
It is a special case of the so-called Vollhardt-W\"olfle Ward identity
proved from a diagrammatic perturbation expansion for the
two-particle vertex $\Lambda$.\cite{Vollhardt80b}

To exploit fully the simplifications brought in by the configurational
averaging we have to switch to a basis set consisting of Bloch waves. It
is a natural basis for translationally invariant lattice systems. The
Bloch waves can be introduced through the Wannier states as follows
\begin{equation}\label{eq:Bloch-states}
\left\langle\mathbf{k}|n\right\rangle =  e^{-i\mathbf{k} \cdot
\mathbf{R}_n}\ ,
\end{equation}
where  $\mathbf{k}$ is quasiparticle momentum labelling the Bloch
waves. As far as we stay in finite volumes, the completeness of the basis
and the norm of any state vector read
\begin{align}\label{eq:completeness-Bloch}
\frac 1N\ \sum_{\mathbf{k}} \left| \mathbf{k} \right\rangle\left\langle
\mathbf{k} \right| = \widehat{1}\ , & \quad ||\psi||^2 \equiv \frac 1N\
\sum_{\mathbf{k}} \left\langle\psi\left| \mathbf{k} \right\rangle
\left\langle \mathbf{k} \right| \psi \right\rangle
\end{align}
with $N$ being the number of sites in the given volume. The only difference
compared to relations \eqref{eq:completeness-Wannier} is the normalization
factor of the basis vectors, $\left\langle \mathbf{k}|\mathbf{k}'
\right\rangle = N\delta_{\mathbf{k},\mathbf{k}'}$. It says that the Bloch
waves are normalized to the volume, i.~e., their norm increases linearly
with the volume. This normalization is chosen so that the thermodynamic
limit is well defined. There the completeness relation comes over to a
spectral integral
\begin{equation}
\frac 1N\ \sum_{\mathbf{k}} \left| \mathbf{k} \right\rangle\left\langle
\mathbf{k} \right|  \xrightarrow[N\to\infty]{} \frac1{(2\pi)^d}\int d^{d}k
\left| \mathbf{k} \right\rangle\left\langle \mathbf{k} \right| =
\widehat{1}
\end{equation}
with generalized orthonormality $\left\langle \mathbf{k}|\mathbf{k}'
\right\rangle = \delta(\mathbf{k}-\mathbf{k}')$, where $\delta$ is the
Dirac delta function.

In the Bloch-wave
representation the one-particle Green function is diagonal
\begin{multline}   \label{eq:av_1PP}
G({\bf k},z)=\frac 1{z-\epsilon({\bf k})-\Sigma({\bf k},z)}\\ =\frac
1N\sum_{ij}e^{-i{\bf k}\cdot({\bf R}_i-{\bf R}_j)}
\left\langle G_{ij}(z)\right\rangle_{av}
\end{multline}
and the two-particle Green function depends on only three momenta as the
total momentum is conserved. The notation for the momentum-dependent
averaged two-particle Green function we use throughout the rest of the
paper reads
\begin{multline}\label{eq:2P_momentum}
G^{(2)}_{{\bf k}{\bf k}'}(z_1,z_2;{\bf q})= \frac
1{N^2}\sum_{ijkl}e^{-i({\bf k} + {\bf q})\cdot{\bf R}_i} e^{i({\bf k}'+
{\bf q})\cdot{\bf R}_j}\\ \times e^{-i{\bf k}'\cdot{\bf R}_k} e^{i{\bf
k}\cdot{\bf R}_l} \left\langle G^{(2)}_{ij,kl}(z_1,z_2)\right\rangle \, .
\end{multline}
The Ward identity \eqref{eq:WI-finite_volume} for the averaged Green
functions in momentum representation is
\begin{equation}
\label{eq:W_BV_momentum} \frac 1N\sum_{{\bf k}'}G^{(2)}_{{\bf k},{\bf
k}'}(z_1,z_2;{\bf 0}) = - \frac 1{\Delta z} \left[ G({\bf k},z_1)-G({\bf
k},z_2)\right]
\end{equation}
where $\Delta z=z_1-z_2$.

Notice that in finite volumes one cannot straightforwardly distinguish
between diffusive electron states, describing a particle ``smeared''
throughout the whole sample, and localized states with exponential tails
confined to a finite subvolume. Such a distinction is meaningful only in
the thermodynamic limit. There the diffusive states have a character of
Bloch waves, i.e., they describe particle {\itshape fluxes}. On the other
hand the localized states contain just a single particle localized in a
finite volume. Intuitively, technical difficulties can arise if a physical
process involves both types of eigenstates, localized and extended. It is
the case of the Anderson localization transition where extended states go
over to localized ones.

\section{Diffusion pole and causality}\label{sec:causality}

Since the averaged two-particle resolvent is not a simple product as for
single configurations, Eq.~\eqref{eq:resolvent-operator}, but rather
fulfills the Bethe-Salpeter equation \eqref{eq:BS-direct}, it is not so
apparent how the Ward identity \eqref{eq:W_BV_momentum} is related with
particle conservation. It can be demonstrated at best via diffusion. In
the weak disorder limit the motion of electrons in a random medium is
expected to be diffusive, i.e., the Fourier components of non-equilibrium
particle-density variations die out according to the exponential law
\begin{equation}
\delta n(t,\mathbf{q})=\delta n(0,\mathbf{q})\,e^{-Dq^2t}\,,\quad t>0\,.
\end{equation}
In order to reproduce such a diffusive behavior the electron-hole
correlation function $\Phi^{AR}_{E_F}(\mathbf{q},\omega)=N^{-2} \sum_{{\bf
k}{\bf k}'} G^{(2)}_{{\bf k}{\bf k}'}(E_F + \omega + i0^+,E_F - i0^+; {\bf
q})$ has to display the so-called diffusion pole,\cite{Janis03a}
\begin{equation}
  \label{eq:Phi_diffusion}
  \Phi^{AR}_{E_F}({\bf q},\omega)\approx  \frac {2\pi n_F}{-i\omega +
    Dq^2} \ .
\end{equation}
Here $D$ stands for the diffusion constant and $n_F$ is the electron
density of states at the Fermi energy $E_F$.  The diffusion pole exists
only if the Ward identity \eqref{eq:W_BV_momentum} is fulfilled by the
averaged one- and two-particle Green functions.\cite{Janis03a}

Unfortunately, the correlation function $\Phi$ does not obey an equation of
motion, so it is difficult to use this realization of the diffusion pole
to control conservation laws in approximate theories. It can, however, be
shown that the same pole must appear in the two-particle irreducible
function $\Lambda$. To demonstrate that we analyze the two-particle
functions in more details. It is convenient to introduce a
vertex $\Gamma$ that is just the two-particle Green function $G^{(2)}$
with uncorrelated part subtracted,
\begin{multline}
G^{(2)}_{{\bf k}{\bf k}'}(z_+,z_-;{\bf q}) = G({\bf k},z_+) G({\bf k} +
\mathbf{q},z_-) \left[\delta({\bf k} - {\bf k}')\right. \\ \left. +
\Gamma_{{\bf k}{\bf k}'}(z_+,z_-;{\bf q}) G({\bf k}',z_+) G({\bf k}' +
\mathbf{q},z_-)\right] .
\end{multline}
In analogy to the averaged two-particle resolvent, the two-particle vertex
obeys a Bethe-Salpeter equation. Once we go beyond local approximations,
an example of which is the mean-field coherent-potential approximation
(CPA), there are three Bethe-Salpeter equations for the vertex $\Gamma$ we
can construct. The three Bethe-Salpeter equations are related to three
different ways how to define a two-particle irreducibility.  Equivalently
they correspond to three topologically distinct two-particle
processes, the electron-hole and the electron-electron (hole-hole)
scatterings and the one-electron self-corrections.\cite{Janis01b}

Since only the off-diagonal one-electron propagators $\bar{G}(\mathbf{k},z)
\equiv G(\mathbf{k},z) - N^{-1}\sum_\mathbf{k}G(\mathbf{k},z)$ are
relevant for distinguishing various representations of the full vertex we
use Bethe-Salpeter equations with just these propagators. The most
important Bethe-Salpeter equation is that from the electron-hole
scattering channel that reads
\begin{subequations}\label{eq:BS}
\begin{multline}\label{eq:BS-eh}
\Gamma_{\mathbf{k}\mathbf{k}'}(\mathbf{q}) =
\bar{\Lambda}^{eh}_{\mathbf{k}\mathbf{k}'}(\mathbf{q}) + \frac
1N\sum_{\mathbf{k}''}
\bar{\Lambda}^{eh}_{\mathbf{k}\mathbf{k}''}(\mathbf{q}) \\ \times
\bar{G}_+(\mathbf{k}'') \bar{G}_-(\mathbf{k}'' + \mathbf{q})
\Gamma_{\mathbf{k}''\mathbf{k}'}(\mathbf{q}) \ .
\end{multline}
We added the superscript $eh$ to the irreducible vertex $\Lambda$ in order
to distinguish it from the other two irreducible channels. Notice that the
full vertex $\Gamma$ must remain channel independent. For the sake of
simplicity we suppressed energy variables. The subscript $\pm$ at the
one-electron propagators refers to the first and second energy variable in
the corresponding two-particle vertex, respectively.

Besides the electron-hole channel, Eq.~\eqref{eq:BS-eh}, it is for our
purposes sufficient to use only one out of the two remaining
Bethe-Salpeter equations --- the electron-electron (hole-hole) channel,
\begin{multline}\label{eq:BS-ee}
\Gamma_{\mathbf{k}\mathbf{k}'}(\mathbf{q}) =
\bar{\Lambda}^{ee}_{\mathbf{k}\mathbf{k}'}(\mathbf{q}) + \frac
1N\sum_{\mathbf{k}''}
\bar{\Lambda}^{ee}_{\mathbf{k}\mathbf{k}''}(\mathbf{q} + \mathbf{k}'  -
\mathbf{k}'')\\ \times \bar{G}_+(\mathbf{k}'') \bar{G}_-(\mathbf{Q}
- \mathbf{k}'') \Gamma_{\mathbf{k}''\mathbf{k}'}(\mathbf{q}+ \mathbf{k} -
\mathbf{k}'')\ ,
\end{multline}
\end{subequations}
where $\mathbf{Q}=\mathbf{k} + \mathbf{k}' + \mathbf{q}$.

The two-particle irreducible vertices play the role of the two-particle
self-energy and are to be determined from e.~g. a diagrammatic expansion.
The best way to access them is to use the so-called parquet
approach.\cite{Janis01b} In it we introduce a completely irreducible
(irreducible in all distinct channels) two-particle vertex $\mathcal{I}$
and use topological nonequivalence of different scattering channels. When
we distinguish only the electron-hole and electron-electron channels we
can write the parquet equation as\cite{Janis01b}
\begin{equation}\label{eq:parquet-equation}
\Gamma_{\mathbf{k}\mathbf{k}'}(\mathbf{q}) =
\bar{\Lambda}^{eh}_{\mathbf{k}\mathbf{k}'}(\mathbf{q})\\ +
\bar{\Lambda}^{ee}_{\mathbf{k}\mathbf{k}'}(\mathbf{q}) -
\mathcal{I}_{\mathbf{k}\mathbf{k}'}(\mathbf{q})\, .
\end{equation}
The minus sign at the completely irreducible vertex $\mathcal{I}$
compensates for the same contributions in the electron-hole and the
electron-electron irreducible vertices.  Notice that all local two-particle
contributions belong to the completely irreducible vertex $\mathcal{I}$.

If the physical system under investigation is invariant with respect to
time reversal (no magnetic field), the two-particle vertices obey the
electron-hole symmetry expressed in a relation
$\Gamma_{\mathbf{k}\mathbf{k}'}(\mathbf{q}) =
\Gamma_{\mathbf{k}\mathbf{k}'}(-\mathbf{q}-\mathbf{k}- \mathbf{k}')$ for
the full vertex and in
$\bar{\Lambda}^{ee}_{\mathbf{k}\mathbf{k}'}(\mathbf{q})=
\bar{\Lambda}^{eh}_{\mathbf{k}\mathbf{k}'}(-\mathbf{q}-\mathbf{k}-
\mathbf{k}')$ for the irreducible ones. As a consequence of the
parquet equation \eqref{eq:parquet-equation} the diffusion pole from the
full vertex $\Gamma$ must appear at least in one of the irreducible
functions $\Lambda$. However, due to the electron-hole symmetry, both
vertices $\Lambda^{eh}$ and $\Lambda^{ee}$ display the same analytic
behavior and hence must contain the same singularity, the diffusion pole.
As discussed above, the existence of the diffusion pole in the
two-particle irreducible vertices is a consequence of conservation
laws applied to averaged systems via the Ward identity. Since the vertex
functions obey equations of motion, the existence of the diffusion pole in
them can be used to check conserving character of approximate
treatments.

Having a pole in the electron-hole irreducible vertex $\Lambda^{eh}$,
however, may have unexpected implications. The electron-hole irreducible
vertex is bound with the self-energy  via the Ward identity. The singular
behavior of the electron-hole irreducible vertex could then be transferred
onto the self-energy.  This can be made transparent by evaluating the
function
\begin{equation} \label{eq:SE_difference}
\Delta W(\omega) = \frac{1}N \sum_{\mathbf{k}} [\Sigma(\mathbf{k},E-\omega
+ i0^+) - \Sigma(\mathbf{k},E +\omega + i0^+)]
\end{equation}
in such a way that the self-energies are expressed in terms of the
two-particle vertex $\Lambda^{eh}$ according to the Fourier transform of
Eq.~\eqref{eq:WI-VW}. Doing so we finally find out\cite{Janis03b} that,
besides an analytic part, the function $\Delta W$ contains also a
singular term of the form
\begin{multline}\label{eq:nonanalyticity}
\Delta W^{sing}_d(\omega)  \thickapprox K \lambda n_F^2 \\ \times
\begin{cases}
 \frac 1{\omega} \left|\frac{\omega}{Dk_F^2}\right|^{d/2}\
 & \text{for $d\neq4l$},\\
\frac 1{\omega} \left|\frac{\omega}{Dk_F^2}\right|^{d/2}\  \ln\left|\frac
{Dk_F^2}\omega\right| & \text{for $d=4l$},
\end{cases}
\end{multline}
where $K$ is a dimensionless constant, $k_F$ stands for the Fermi momentum
and $d$ denotes spatial dimension. The non-analyticity of the
self-energy, and therefore of the one-electron Green function, is
unacceptable as it contradicts another  very fundamental property ---
causality. Causality is a consequence of self-adjointness of the
Hamiltonian, that is of reality of eigenenergies. It  cannot be given up
in any physically meaningful treatment. We are thus led to a surprising
conclusion that the Ward identity \eqref{eq:WI-VW} cannot survive to the
thermodynamic limit of configurationally averaged Green
functions. At least for real-energy differences denoted $\omega$ in
Eq.~\eqref{eq:SE_difference} and \eqref{eq:nonanalyticity}.
This indicates that there must be a flaw in the seemingly
straightforward extension of the Ward identity from finite to infinite
configurationally averaged systems.

\section{Approximate analytic treatments}\label{sec:approximate}

It is impossible to resolve rigorously the incompatibility of
probability conservation and causality in infinite, configurationally
averaged systems, since exact solutions are not provided. We hence have to
resort to approximate treatments to trace down the problem and clues for
a possible resolution.
 
It is natural to start with a mean-field solution for disordered electron
systems, being the exact solution of the problem in infinite
spatial dimensions. This limit amounts to the coherent-potential
approximation. We have explicit expressions for all quantities in this
approximation. The self-energy is determined from a Soven equation that
can be cast into
\begin{subequations}\label{eq:CPA_equations}
 \begin{equation}\label{eq:CPA_1}
 G(z)=\left\langle\left[G^{-1}(z) + \Sigma(z) -
V_i\right]^{-1}\right\rangle_{av}
 \end{equation}
 where $G(z)=N^{-1}\sum_\mathbf{k}G(\mathbf{k},z)$. The two-particle
 irreducible vertex then is
\begin{widetext} \begin{multline}\label{eq:CPA_2}
    \lambda(z_+,z_-)=\frac 1{G(z_+)G(z_-)}\left[1- {\left\langle \frac 1{
            1+\left(\Sigma(z_+)-V_i\right)G(z_+)}\frac 1{ 1+\left(
              \Sigma(z_-)-V_i\right)G(z_-)}\right\rangle_{av}}^{-1}
    \right] = \frac {\Sigma(z_+) - \Sigma(z_-)} {G(z_+) - G(z_-)}\ .
 \end{multline}\end{widetext}
\end{subequations}
The self-energy from the Soven equation \eqref{eq:CPA_1} can be proven
analytic and  the last equality in Eq.~\eqref{eq:CPA_2} expresses the Ward
identity. The conservation law is fulfilled and apparently does not
contradict causality.

The CPA is indeed a consistent theory, but only if we keep strictly to
local quantities.  To remain consistent at the two-particle level we
should hence consider only the local vertex being $\gamma(z_+,z_-) =
\lambda(z_+,z_-)/(1 - \lambda(z_+,z_-) G(z_+)G(z_-))$. This local vertex,
however, does not contain the diffusion pole. Velick\'y derived in
Ref.~\onlinecite{Velicky69} the full two-particle vertex in the CPA
in a form
\begin{equation}\label{eq:full_vert_CPA}
\Gamma_{\mathbf{k}\mathbf{k}'}(\mathbf{q})=\frac{\lambda}{
  1-\lambda\,\chi(\mathbf{q})}
\end{equation}
where $\chi$ is a two-electron bubble
\begin{equation}\label{eq:bubble}
\chi(\mathbf{q}) = \frac 1N\sum_{\mathbf{k}}
G_+(\mathbf{k}) G_-(\mathbf{k} + \mathbf{q})\,.
\end{equation}
The CPA nonlocal two-particle vertex, Eq.~\eqref{eq:full_vert_CPA},
contains the diffusion pole precisely as in Eq.~\eqref{eq:Phi_diffusion},
see Ref.~\onlinecite{Janis03a} for explicit calculation. Even at this
level the CPA seems to be consistent. The CPA does not treat, however, the
electron-hole and the electron-electron vertex equivalently. We have
$\Lambda^{eh}=\lambda$, whereby there is no expression for the
electron-electron irreducible vertex $\Lambda^{ee}$. The electron-hole
symmetry for two-particle vertices is therefore broken in the mean-field
solution. As a consequence of the violated electron-hole symmetry at the
two-particle level we have no sign of localized states in the CPA.

To include localized states and to approach the Anderson metal-insulator
transition we have to go beyond local approximations. Already the notion
of weak localization needs a sum of infinite-many ``crossed'' two-particle
diagrams (ladders from the electron-electron channel) leading to a
singular electron-hole irreducible vertex. The importance of the
electron-hole symmetry at the two-particle level and the diffusion pole in
the electron-hole irreducible vertex was stressed and exploited by
Vollhardt and W\"olfle in their theory of Anderson
localization.\cite{Vollhardt80b}
 
Theory of Vollhardt and W\"olfle is a self-consistent approximation for the
electron-hole and electron-electron irreducible vertices. Due to the
complete electron-hole symmetry we obtain a nonlinear equation for a
generic vertex. This vertex displays the diffusion pole of
Eq.~\eqref{eq:Phi_diffusion}. Further on, the Ward identity
\eqref{eq:WI-VW} is assumed to hold. In this situation the two-particle
irreducible vertex in the hydrodynamic limit ($q\to 0$) contains only a
single free adjustable parameter, the diffusion constant $D$. Anderson
localization is signalled in this approach by vanishing of the diffusion
constant $D$.

A diagrammatic expansion for two-particle irreducible functions is formally
used to select an appropriate approximation in the Vollhardt-W\"olfle
theory. This theory, however, does not provide a similarly closed
expression for the one-electron self-energy. Since the spectral behavior
of the self-energy is not decisive for the existence of the Anderson
localization transition, the self-energy in the one-electron propagators
is taken from the weak-scattering limit, i.~e. from the Born
approximation. The Ward identity is used only to recover the diffusion
pole, but not to determine the self-energy consistently so that it be
compatible with the chosen approximate vertex function. It is then evident
that we cannot check compatibility of the Ward identity with causality
within the Vollhardt-W\"olfle theory. Due to Eq.~\eqref{eq:nonanalyticity}
we cannot expect that a causal self-energy could be found to the
approximate vertex resulting from the Vollhardt-W\"olfle approach.
 
We can use another route to trace conservation laws in configurationally
averaged infinite systems. Since the CPA obeys all restrictive conditions
except for the two-particle electron-hole symmetry, we can try to correct
this approximation just by implementing this missed feature. This can
actually be achieved within the parquet approach providing a framework to
sum systematically nonlocal vertex corrections to the CPA irreducible
vertex.\cite{Janis01b} We  showed recently how to solve the parquet
equations in electron-hole symmetric theories asymptotically exactly in
the limit of high spatial dimensions.\cite{Janis04a} The solution for the
full two-particle vertex resembles that from the CPA, but the
electron-hole symmetry is manifestly present. We obtain\cite{Janis04a}
\begin{multline}\label{eq:vertex-full}
\Gamma_{\mathbf{k}\mathbf{k}'}(\mathbf{q}) = \gamma \\ + \Lambda_0
\left[\frac{\bar{\Lambda}_0\bar{\chi}(\mathbf{q})} {1 - \Lambda_0
\chi(\mathbf{q})} + \frac{\bar{\Lambda}_0\bar{\chi}(\mathbf{k} +
\mathbf{k}' + \mathbf{q})} {1 - \Lambda_0 \chi(\mathbf{k} + \mathbf{k}' +
\mathbf{q})}\right]\
\end{multline}
where we used the local CPA vertex $\gamma$ and denoted $\Lambda_0 =
\bar{\Lambda}_0/(1 + \bar{\Lambda}_0G_+ G_-)$ and $\bar{\chi}(\mathbf{q})
= \chi(\mathbf{q}) - G_+ G_-$. The only local parameter to be determined
from the parquet equation is $\bar{\Lambda}_0$. It fulfills an equation
\begin{equation}\label{eq:parquet-local}
\bar{\Lambda}_0 = \gamma + \bar{\Lambda}_0 \frac 1N \sum_{\mathbf{q}} \frac
{\bar{\Lambda}_0 \bar{\chi}(\mathbf{q})}{1 - \bar{\Lambda}_0
\bar{\chi}(\mathbf{q})} \ .
\end{equation}
It is clear that the CPA vertex, Eq.~\eqref{eq:full_vert_CPA}, is
reproduced from Eq.~\eqref{eq:vertex-full} if we put
$\bar{\Lambda}_0=\gamma$ in it and if the last term on its right-hand side
is neglected. It is just the neglected term that restores the two-particle
electron-hole symmetry. On the other hand, it is the same term that makes
it difficult to comply with the Ward identity. The vertex $\Gamma$ from
Eq.~\eqref{eq:vertex-full} does not obey a Bethe-Salpeter equation. It
is a sum of two solutions of the Bethe-Salpeter equations in the
electron-hole and the electron-electron channels. It is hence unclear how
to choose the self-energy so that it would fit the Ward identity.

To restore the diffusive behavior of the full vertex we have to use the
Ward identity for a specific combination of energies and determine the
self-energy from the irreducible vertex $\Lambda_0$. We use\cite{Janis04a}
\begin{equation}\label{eq:self-energy}
\Im\Sigma(E+i\eta) = \Lambda_0(E+i\eta, E-i\eta) \Im G(E+i\eta) \ .
\end{equation}
to resolve the imaginary part of the self-energy. The real part of the
self-energy is then calculated from the Kramers-Kronig relation
 \begin{equation}
  \label{eq:SE_KK1}
  \Re \Sigma(E+i\eta) = \Sigma_\infty + P\int_{-\infty}^{\infty}
   \frac{d E'}{\pi} \ \frac{\Im\Sigma(E'+i\eta)}{E'-E}\ .
\end{equation}
In this way we construct a consistent approximation with the two-particle
electron-hole symmetry and with a causal self-energy that satisfies the
Ward identity in a maximally possible way. Since the Ward identity was
used in Eq.~\eqref{eq:self-energy} only for imaginary energy difference in
the vertex function, this solution is not in conflict with
Eq.~\eqref{eq:nonanalyticity} where the energy difference is real.

The Ward identity does not hold in this solution for real energy
differences. As a consequence the diffusion pole from
Eq.~\eqref{eq:Phi_diffusion} modifies to
\begin{equation}
  \label{eq:Phi-diffusion-modified}   \Phi^{AR}_{E_F}({\bf
q},\omega)\approx  \frac {2\pi n_F}{-i A(\omega) \omega + D(\omega)q^2} \ .
\end{equation}
The weight of the pole is no longer one, $A(0)\ge 1$. The
quasistatic and hydrodynamic limit of the vertex function is then
governed by two parameters, the diffusion constant $D(0)$ and the weight
of the diffusion pole (wave-function normalization) $A(0)^{-1}$. The
Anderson localization transition is signalled here by vanishing of the
weight of the diffusion pole $A(0)^{-1}\to 0$ accompanied by vanishing of
the effective diffusion constant $D'=D(0)/A(0)\to 0$.

\section{Problems with a quantitative description of localized states in
translationally invariant systems}\label{sec:Anderson}

Our analysis disclosed a surprising fact that it is impossible to describe
consistently disordered electron systems with averaged Green functions so
that conservation of particle number would hold in the state space of
Bloch waves, the eigenstates of the pure solids. Since the probability
density is not conserved in the description with averaged functions, we
have to ask whereto can particles vanish?

To answer this question and to understand deviations from the Ward identity
in averaged infinite systems, we recall that there are two very
different types of states in the thermodynamic limit --- extended
and localized. The former are normalized to volume. Their
normalization factor increases linearly with increasing volume. The latter
are normalized to unity. Their normalization constant is effectively
independent of the volume of large samples. This difference may seem to be
a formal issue, but in fact it poses a principal obstacle for describing
both types of states within one theoretical framework.

Each configuration of the random potential in a disordered system has its
own representation space. It is the Hilbert space spanned over the
eigenstates of the Hamiltonian with given values of the random
potential. We can generally characterize the representation space via a
decomposition of unity operator
\begin{equation}\label{eq:unity-mixed}
\widehat{1} = \frac 1{N_{ext}}\sum_\mathbf{x}^{N_{ext}}
|\mathbf{x}\rangle\langle \mathbf{x}| + \sum_\nu^{N_{loc}}
|\nu\rangle\langle \nu |
\end{equation}
where we distinguished extended eigenstates $|\mathbf{x}\rangle$ and
localized ones $|\nu\rangle$. The total number of eigenstates is $N_{ext}
+ N_{loc} = N$. To determine the eigevalues of the configurationally
dependent Hamiltonian we have to specify boundary conditions for the
eigenvalue problem. We standardly use periodic boundary conditions for the
extended states and asymptotic vanishing of the localized states. It is
evident that we cannot describe localized states with periodic boundary
conditions, unless they are periodically repeated as are e.~g. Wannier
states. Configurationally dependent localized states are confined to a
specific finite part of the space and do not replicate themselves
periodically, since one configuration is typically not translationally
invariant.

We cannot effectively work with individual configurations, since we know
neither the extended nor the localized eigenstates. Moreover, the ratio
$N_{loc}/N_{ext}$ varies from configuration to configuration. In the
thermodynamic limit we are able to deal only with extended or periodically
distributed localized states. We standardly use the Hilbert space spanned
over the Bloch waves and assume that all relevant configurations of the
random Hamiltonian can be accommodated in this space. There are no
problems with delocalized eigenstates, since we have $\langle
\mathbf{x}|\mathbf{k}\rangle/\langle\mathbf{k}|\mathbf{k}\rangle = O(1)$.
However, the localized, translationally noninvariant states fall out from
this Hilbert space, since $\langle
\nu|\mathbf{k}\rangle/\langle\mathbf{k}|\mathbf{k}\rangle =
O(N^{-1})$. Hence, if we have macroscopically many configurations with
localized eigenstates, we observe deviations from the Ward identity for
the averaged Green functions and probability will no longer be conserved
in the space of asymptotically free states. Mathematically it means that
the number of extended eigenstates $N_{ext}$ is configurationally dependent
and we cannot interchange the configurational averaging with the summation
over the intermediate states in Eq.~\eqref{eq:averaging-commutativity}.
 
Evasion of localized states from the space of extended waves is a severe
drawback of the description, since we loose the normalization of the
density of states. To repair this we replace the missing localized states
with extended ones so that we keep the number of available states always
equal the number of the lattice sites $N$. Thereby we believe that the
differences between the \textit{exact unknown localized} and the
\textit{supplemented known Bloch waves} can be represented by corrections
via the self-energy and other irreducible vertex functions. We recover the
sum rule for the density of states, but the Ward identity at best only in
the quasistatic limit $\omega\to 0$. This is the price we pay for our
inability to describe the localized states in translationally invariant
systems exactly.

\section{Conclusions}\label{sec:conclusions}
 
We discussed in this paper applicability of conservation laws in disordered
itinerant noninteracting systems. We showed that formal manipulations
leading to conservation of the norm of the wave function are justifiable
only for individual configurations of the random potential in finite
volumes. Extensions of the conservation laws via Ward identities to
infinite systems demand a caution when dealing with infinite sums. In the
thermodynamic limit we have to distinguish proper states from the Hilbert
space with a finite norm (localized states) and generalized states with
the norm (or the normalization constant) proportional to the volume
(extended states). If we stay within individual configurations of the
random potential we can well separate these two types of states and rely
on the derivation of the Ward identities.

Problems arise, however, when we start to average over configurations to
restore translational invariance. A generalization of the Ward identity to
the averaged Green functions faces principal obstacles, when both types of
states, localized and extended, are present. An extension of the Ward
identity to the averaged functions can work when eigenstates of all
relevant configurations are from the same Hilbert space. Based on the
observed behavior of averaged systems we conclude that, whenever localized
states become macroscopically relevant, we are unable to find the proper
Hilbert space to represent the localized eigenstates exactly. We have at
our disposal practically only the Hilbert space spanned over the Bloch
waves. The localized states are in the thermodynamic limit orthogonal to
Bloch waves,
$\langle\nu|\mathbf{k}\rangle/\langle\mathbf{k}|\mathbf{k}\rangle =
O(N^{-1})$, and hence elusive to a quantitative description within the
available Hilbert space. Since the space of Bloch waves is incomplete,
probability is no longer conserved, when macroscopic portion of
configurations contain localized states. Our standard assumption that
configurational averaging does not influence the representation space and
can be represented as averaging over rotations in the given Hilbert space
is hence incorrect when quantum coherence between spatially distinct
scattering events is taken into account.

Research on this problem was carried out within a project AVOZ1-010-914
of the Academy of Sciences of the Czech Republic and supported in part by
Grant No. 202/04/1055 of the Grant Agency of the Czech Republic.

\end{document}